# Traffic flow optimization using a quantum annealer

Florian Neukart[*1], David Von Dollen[1], Gabriele Compostella[2], Christian Seidel[2], Sheir Yarkoni[3], and Bob Parney[3]

[1]Volkswagen Group of America, San Francisco, USA
[2]Volkswagen Data:Lab, Munich, Germany
[3]D-Wave Systems, Inc., Burnaby, Canada

**Abstract**

Quantum annealing algorithms belong to the class of meta-heuristic tools, applicable for solving binary optimization problems. Hardware implementations of quantum annealing, such as the quantum processing units (QPUs) produced by D-Wave Systems, have been subject to multiple analyses in research, with the aim of characterizing the technology's usefulness for optimization and sampling tasks. In this paper, we present a real-world application that uses quantum technologies. Specifically, we show how to map certain parts of a real-world traffic flow optimization problem to be suitable for quantum annealing. We show that time-critical optimization tasks, such as continuous redistribution of position data for cars in dense road networks, are suitable candidates for quantum computing. Due to the limited size and connectivity of current-generation D-Wave QPUs, we use a hybrid quantum and classical approach to solve the traffic flow problem.

[*]Corresponding author: `florian.neukart@vw.com`

# 1 Introduction

Quantum annealing technologies such as the quantum processing units (QPUs) made by D-Wave Systems are designed to solve complex combinatorial optimization problems. It has been shown in literature how to use these QPUs to perform both complex sampling and optimization tasks, and how the properties of the quantum bits (qubits) play a role in the computation of solutions [4, 6, 8–13]. The QPU is designed to solve quadratic unconstrained binary optimization (QUBO) problems, where each qubit represents a variable, and couplers between qubits represent the costs associated with qubit pairs. The QPU is a physical implementation of an undirected graph with qubits as vertices and couplers as edges between them. The functional form of the QUBO that the QPU is designed to minimize is:

$$\text{Obj}(x, Q) = x^T \cdot Q \cdot x, \qquad (1)$$

where $x$ is a vector of binary variables of size $N$, and $Q$ is an $N \times N$ real-valued matrix describing the relationship between the variables. Given the matrix $Q$, finding binary variable assignments to minimize the objective function in Equation 1 is equivalent to minimizing an Ising model, a known NP-hard problem [7].

In this paper, we will introduce the traffic flow optimization problem. We start with the T-Drive trajectory dataset[1] of cars' GPS coordinates, and develop a workflow to mimic a system that aims to optimize traffic flow in real time. We show how to transform key elements of the problem to QUBO form, for optimization on the D-Wave system (including both the machine and software tools that use it). We treat the D-Wave system as an optimizer, and show that it is possible to integrate D-Wave QPU calls into a workflow that resembles a real-world application. The method presented here is a novel approach to mapping this real-world problem onto a quantum computer.

# 2 Formulation of the traffic flow problem

The objective of the traffic flow optimization problem is to minimize the time for a given set of cars to travel between their individual sources and destinations, by minimizing total congestion over all road segments. Congestion on an individual segment is determined by a quadratic function of the number of cars traversing it in a specific time interval. To ensure reproducibility, we used the publicly available T-Drive trajectory dataset containing trajectories of 10,357 taxis recorded over one week. The dataset features 15 million data points, and the total distance of the trajectories makes up about 9 million kilometers [14–16]. We required every car to transmit its GPS coordinates in intervals of 1 to 5 seconds. Because not all cars in the dataset provide transmission data at this rate, we enriched the dataset by interpolating between GPS points. We split the problem into a step-by-step workflow, outlined below. "Classical" refers to calculations on classical machines, and "quantum" refers to calculation on the D-Wave system:

---

[1]This open source dataset provided by Microsoft can be found here:
https://www.microsoft.com/en-us/research/publication/t-drive-trajectory-data-sample/

1. Classical: Pre-process map and GPS data.

2. Classical: Identify areas where traffic congestion occurs.

3. Classical: Determine spatially and temporally valid alternative routes for each car in the dataset, if possible.

4. Classical: Formulate the minimization problem as a QUBO (to minimize congestion in road segments on overlapping routes).

5. Hybrid Quantum/Classical: Find a solution that reduces congestion among route assignments in the whole traffic graph.

6. Classical: Redistribute the cars based on the results.

7. Iterate over steps 2 to 6 until no traffic congestion is identified.

A visualization of the input graph is shown in Figure 1. This visualization was generated using the OSMnx API, which is based on OpenStreetMap and allows for retrieving, constructing, analyzing, and visualizing street networks from OpenStreetMap [1].

## 2.1 Determination of alternate routes

To illustrate how we formulate the problem, we focus on a subset of the T-Drive dataset. Of the 10,357 cars in the dataset, we select 418 of those that are traveling to or from the city center and the Beijing airport. In this specific scenario, the goal was to maximize traffic flow by redirecting a subset of the 418 cars to alternative routes such that the number of intersecting road segments is minimized. For this, optimizing over all cars simultaneously is required, which means that any redistribution of cars that resolves the original congestion must not cause a traffic jam anywhere else in the map. We used the OSMnx package to split the map of Beijing into segments and nodes, and assign a unique ID to each. Our procedure can be summarized as follows:

1. Extract the road graph from the Beijing city map using OSMnx. This returns lists of segments and nodes with IDs. Nodes represent connections between segments, and segments are edges connecting the nodes, representing the streets (Figure 1).

2. Map the T-Drive trajectory dataset cars' GPS coordinates onto street segments in the graph, to determine the routes taken by the cars.

3. For each car, and each source and destination node, we extract all simple paths from source to destination, and obtain 3 candidate alternative routes[2]. We use these 3 candidates as proposed alternative routes to redistribute traffic.

---

[2] A simple path can traverse several nodes from source to destination, but without returning to nodes which were already visited (no cycles). Several thousands of simple paths from source to destination (per car) may exist. We selected two simple paths that are most dissimilar to the original route, and to each other, and proposed these as alternates, along with the original route. To do this we used the Jaccard similarity index.

## 2.2 Formulating the traffic flow optimization in QUBO form

The definition of variables for the QUBO (Equation 1) requires some classical pre-processing on the input. In rare cases it may not be possible to switch a car to different route. For example, if there is no intersection or ramp near the car, it will not be considered for rerouting and will remain on its original path. Nevertheless, this car will still affect possible routings of other cars, so it is included in the QUBO. Figure 2 shows an example with road segments assigned to a car, as it is used in our workflow.

To optimize the traffic flow, we minimize the number of overlapping segments between assigned routes for each car. Thus, we formulate the optimization problem as follows: "Given 3 possible routes per car, which assignment of cars to routes minimizes the overall congestion on all road segments?" We require that every car be assigned one of the 3 possible routes, while simultaneously minimizing total congestion over all assigned routes. It is important to emphasize that in this example each car was proposed 3 possible alternative routes — not the same set of 3 routes for all cars. This not need be the case in general; cars can have many possible routes. For simplicity we take (maximum) 3 routes per car, because the mathematical description of the problem is identical regardless of the number of routes.

For every possible assignment of car to route, we define a binary variable $q_{ij}$ representing car $i$ taking route $j$. Because each car can only occupy one route at a time, exactly one variable per car must be true in the minimum of the QUBO. We define a constraint such that every car is required to take exactly one route. This can be formulated as the following constraint (assuming 3 possible routes):

$$0 = \left(\sum_{j\in\{1,2,3\}} q_{ij} - 1\right)^2 = -q_{i1} - q_{i2} - q_{i3} + 2q_{i1}q_{i2} + 2q_{i2}q_{i3} + 2q_{i1}q_{i3} + 1, \qquad (2)$$

simplified using the binary rule $x^2 = x$. As stated previously, routes are described by lists of street segments ($S$ being the set of all street segments in the graph). Therefore, for all cars $i$, with proposed routes $j \in \{1, 2, 3\}$ with segments $S_j$, which share the street segment $s \in S_j$, the cost of the occupancy of the street segment is given by:

$$\text{cost}(s) = \left(\sum_i \sum_j \sum_{s \in S_j} q_{ij}\right)^2 \qquad (3)$$

The global cost function for the QUBO problem, Obj from Equation 1, can now be simply described by summing the cost functions for each street segment and the constraint from Equation 2:

$$\text{Obj} = \sum_{s\in S} \text{cost}(s) + \lambda \sum_i \left(\sum_j q_{ij} - 1\right)^2. \qquad (4)$$

When summing components of the global cost function, the scaling parameter $\lambda$ must be introduced. This ensures that Equation 2 is satisfied for all cars in the minimum of the QUBO. To find this scaling factor, we find the maximum number of times some car $i$ is present in cost functions of the form Equation 3, and use this value as $\lambda$. This makes the cost of violating Equation 2 greater than the cost of increasing the segment occupancy in every route by 1.

Now the cost function can be formulated as a quadratic, upper-triangular matrix, as required for the QUBO problem. We keep a mapping of binary variable $q_{ij}$ to index in the QUBO matrix $Q$ (as defined in Equation 1), given by $I(i,j)$. These indices are the diagonals of the QUBO matrix. The elements of the matrix are the coefficients of the $q_{ij}$ terms in Equation 4. To find these terms explicitly, whenever two routes $j$ and $j'$ share a street segment $s$:

1. We add a $(+1)$ at diagonal index $I(i,j)$ for every car $i$ proposed with route $j$ containing segment $s$.

2. We add a $(+2)$ for every pair of cars $i_1$ and $i_2$ taking route $j$ containing segment $s$ at the off-diagonal index given by indices $I(i_1,j)$ and $I(i_2,j)$.

We then add the constraints to enforce that every car has only one route, as per Equation 2:

1. For every car $i$ with possible route $j$, we add $(-\lambda)$ to the diagonal of $Q$ given by index $I(i,j)$.

2. For every cross-term arising from Equation 2, we add $(2\lambda)$ to the corresponding off-diagonal term.

A special case occurs if a car is proposed only one route, meaning $q_{ij} = 1$. As stated previously, despite car $i$ being assigned to route $j$, this assignment still affects other cars. This forces the quadratic constraint terms from Equation 3 to be turned into additional linear terms: $2q_{ij}q_{i'j'} \to 2q_{i'j'}$.

Additionally, by keeping a record of which routes every segment appears in, we can remove the redundant constraints, as some routes may overlap in more than one segment.

This results in a QUBO matrix as shown in Figure 3.

## 2.3 Summary of the traffic flow optimization algorithm

Expressed as pseudo-code, the important high-level steps of the traffic flow optimization algorithm are as follows:

1. For each car $i$

    (a) Determine the current route.

2. For each car $i$'s current route:

    (a) Map the source and destination to their nearest nodes in the road graph.

3. For each with source/destination pair:

    (a) Determine all simple paths from source to destination.

(b) Find two alternative paths that are maximally dissimilar to the original route and to each other.

4. For each car $i$, define the set of possible routes needed to form the QUBO.

5. Define the matrix $Q$ with binary variables $q_{ij}$ as described in Section 2.2.

6. Solve the QUBO problem.

7. Update cars with the selected routes.

# 3 D-Wave solvers and architecture

Here, we briefly introduce the solvers and tools provided by D-Wave, to help understand how the problem was solved using the QPU.

## 3.1 Connectivity and topology

The topology of the D-Wave 2X QPU is based on a $C_{12}$ Chimera graph containing 1152 vertices (qubits) and over 3000 edges (couplers). A Chimera graph of size $C_N$ is an $N \times N$ grid of Chimera cells (also called unit tiles or unit cells), each containing a complete bipartite graph of 8 vertices ($K_{4,4}$). Each vertex is connected to its four neighbors inside the cell as well as two neighbors (north/south or east/west) outside the cell: therefore every vertex has degree 6 excluding boundary vertices [5].

The 418-car example used 1254 logical variables to represent the problem. A challenge in this scenario is the restricted connectivity between qubits on a D-Wave QPU, which limits the ability to directly solve arbitrarily-structured problems. When using the D-Wave QPU directly, an interaction between two problem variables can only occur when there is a physical connection (coupler) between the qubits representing these variables. For most problems, the interactions between variables do not match the QPU connectivity. This limitation can be circumvented using minor-embedding, a technique that maps one graph structure to another. The QPU we used has 1135 functional qubits, thus it was not possible to embed the 1254 logical variables on the QPU at once. Therefore, the problem was solved using the hybrid classical/quantum tool `qbsolv` (described in the next section).

## 3.2 The `qbsolv` algorithm

In January 2017, D-Wave Systems open-sourced the software tool `qbsolv` [3] [3]. The purpose of this algorithm is to provide the ability to solve larger QUBO problems, and with higher connectivity, than is currently possible on the QPU. Given a large QUBO input, `qbsolv` partitions the input into important components and then solves the components independently using queries to the QPU. This process iterates (with different components found by Tabu search) until no improvement in the solution is found. The `qbsolv` algorithm can optimize sub-problems using either a classical Tabu solver or via submission to a D-Wave QPU. In this paper, we run `qbsolv` in the hybrid

---

[3] The source code can be found at: `github.com/dwavesystems/qbsolv`

classical/quantum mode of submitting sub-problems to the D-Wave 2X QPU.

The high-level steps performed by `qbsolv` in hybrid mode are as follows:

1. Find the largest clique[4] that can be minor embedded in the QPU topology, or in the full Chimera graph if using the VFYC feature[5]. This one-time operation can be done in advance.

2. Given a QUBO problem, initialize random bit-string representing a solution to the problem.

3. Use a heuristic method to rank nodes according to importance; create a sub-problem that fits on the QPU using the importance ranking.

4. Create sub-problem using the importance order.

5. Solve sub-problem by submitting it to the QPU and update variable states in the bit-string.

6. Iterate steps 3 to 5 until no improvement in the objective function is found.

A full description of how the `qbsolv` algorithm works is detailed in [2].

# 4 Results

The goal of these experiments was to map a real-world problem to a quantum annealing machine, which we have shown. When evaluating the solutions produced by the D-Wave QPU, the focus was on finding good quality solutions within short periods of calculation. To quantify the quality of a solution, we count the number of congested roads after optimization. Keeping in mind that routes are described by sets of road segments, we simply count the number of segments that appear in routes more than a given number of times ($N_{\text{intersections}}$). Here we assume that a segment that appears in more than $N_{\text{intersections}}$ routes will become congested. For this experiment, we chose $N_{\text{intersections}} = 10$.

To evaluate the QUBO formulation of the traffic flow problem, we designed the following experiment: for the 418 car QUBO problem (as presented in Section 2.2), we solved the problem 50 times using `qbsolv`. We also generated 50 random assignments of cars to routes as reference for the results. Intuitively, one would expect random route assignments to spread traffic across the alternative routes, thus reducing the number of congested segments. In Figure 4 we show the distribution of results (measured as the number of congested segments) after running the experiments using `qbsolv` and random assignments.

From the results in Figure 4, we can see that `qbsolv` redistributes the traffic over possible routes in a way that reduces the number of congested roads. This is evident both with respect to random assignment of routes, and also shows improvement over the original assignment of routes. It should be noted that in the original assignment, there was a relatively small number of streets that are heavily occupied (meaning above the $N_{\text{intersections}} = 10$ threshold), as all the cars shared the same route, and that the average occupancy was much higher than $N_{\text{intersections}} = 10$. It is also worth noting that all 50 experiments using `qbsolv` resolved the congestion.

---

[4] A clique is a graph where all nodes are connected to each other.

[5] D-Wave has recently introduced a "virtual full-yield Chimera" (VFYC) solver, which takes the working QPU and simulates the missing qubits and couplers using classical software. This allows for some programs to be standardized across the different QPUs, and within generations of QPUs. This VFYC version of the D-Wave 2X solver was used in our experiments.

Additionally, we measured the performance of `qbsolv` as a function of its run-time. The `qbsolv` source code was compiled and executed on a server in Burnaby, Canada, to minimize the latency between submitting jobs to the QPU and obtaining the results. However, since the QPU used was a shared resource via the cloud, run-time of `qbsolv` varied greatly. Therefore, we consider the run-time of `qbsolv` to be the minimum of the observed run-times, as this represents most faithfully the algorithm, independent of the load on the D-Wave system. This run-time was observed as 22 seconds. There is also no evidence of correlation between the run-time of `qbsolv` and performance (the long run-times are due to waiting in the job submission queue). Given the performance results of `qbsolv`, it is reasonable to assume that a dedicated D-Wave QPU (circumventing the public job submission queue) could be suitable for these kinds of optimization problems. A visual showing the traffic density on the Beijing road graph before (original routes) and after optimization (using `qbsolv`) is shown in Figure 5.

## 5 Conclusions and future work

The currently presented problem is a simplified version of traffic flow, as it incorporates only a limited set of cars, no communication to infrastructure, no other traffic participants, and no other optimization targets except minimization of road congestion. In our future work, we intend to consider all of these parameters, and will also need to consider creative ways of formulating these parameters as part of the QUBO problem. We will continue to focus on solving real-world problems by means of quantum machine learning, quantum simulation, and quantum optimization. Furthermore, we find that these types of real-time optimization problems are well-suited for the D-Wave systems, and the hybrid tools that use them. The more combinatorially complex the problem becomes, the more time would be needed for classical algorithms to consider additional parameters. However, D-Wave QPUs have historically grown in number of qubits from one generation to the next, and given that this trend is likely to continue, it is reasonable to assume that obtaining high-quality solutions quickly using the QPU will be sustainable moving forward. We expect that in future generations of QPUs, we will be able to embed larger problems directly. This will allow us to further leverage the performance of the QPU.

## Acknowledgments

Thanks go to the Volkswagen Group for its support in this exploratory research project. Further thanks go to the team at D-Wave Systems, especially to Murray Thom, Adam Douglass, and Andy Mason.

# References

[1] Geoff Boeing. Osmnx: New methods for acquiring, constructing, analyzing, and visualizing complex street networks. *Computers, Environment and Urban Systems*, 65:126 – 139, 2017.

[2] Michael Booth, Steven P. Reinhardt, and Aidan Roy. Partitioning optimization problems for hybrid classical/quantum execution. https://www.dwavesys.com/resources/publications, 2017.

[3] D-Wave Systems. D-Wave Initiates Open Quantum Software Environment. https://www.dwavesys.com/press-releases/d-wave-initiates-open-quantum-software-environment, Jan 2017.

[4] Vasil S. Denchev, Sergio Boixo, Sergei V. Isakov, Nan Ding, Ryan Babbush, Vadim Smelyanskiy, John Martinis, and Hartmut Neven. What is the computational value of finite-range tunneling? *Phys. Rev. X*, 6:031015, Aug 2016.

[5] James King, Sheir Yarkoni, Mayssam M. Nevisi, Jeremy P. Hilton, and Catherine C. McGeoch. Benchmarking a quantum annealing processor with the time-to-target metric. arXiv:1508.05087, 2015.

[6] Los Alamos National Laboratory. D-Wave Rapid Response. http://www.lanl.gov/projects/national-security-education-center/information-science-technology/dwave/, 2016.

[7] Andrew Lucas. Ising formulations of many np problems. *Frontiers in Physics*, 2:5, 2014.

[8] B. O'Gorman, R. Babbush, A. Perdomo-Ortiz, A. Aspuru-Guzik, and V. Smelyanskiy. Bayesian network structure learning using quantum annealing. *The European Physical Journal Special Topics*, 224(1):163–188, Feb 2015.

[9] A. Perdomo-Ortiz, J. Fluegemann, S. Narasimhan, R. Biswas, and V.N. Smelyanskiy. A quantum annealing approach for fault detection and diagnosis of graph-based systems. *The European Physical Journal Special Topics*, 224(1):131–148, Feb 2015.

[10] Jack Raymond, Sheir Yarkoni, and Evgeny Andriyash. Global warming: Temperature estimation in annealers. *Frontiers in ICT*, 3:23, 2016.

[11] Eleanor G. Rieffel, Davide Venturelli, Bryan O'Gorman, Minh B. Do, Elicia M. Prystay, and Vadim N. Smelyanskiy. A case study in programming a quantum annealer for hard operational planning problems. *Quantum Information Processing*, 14(1):1–36, Jan 2015.

[12] Davide Venturelli, Salvatore Mandrà, Sergey Knysh, Bryan O'Gorman, Rupak Biswas, and Vadim Smelyanskiy. Quantum optimization of fully connected spin glasses. *Phys. Rev. X*, 5:031040, Sep 2015.

[13] Davide Venturelli, Dominic J. J. Marchand, and Galo Rojo. Quantum annealing implementation of job-shop scheduling. arXiv:1506.08479, 2015.

[14] J. Yuan, Y. Zheng, X. Xie, and G. Sun. T-drive: Enhancing driving directions with taxi drivers' intelligence. *IEEE Transactions on Knowledge and Data Engineering*, 25(1):220–232, Jan 2013.

[15] Jing Yuan, Yu Zheng, Xing Xie, and Guangzhong Sun. Driving with knowledge from the physical world. In *Proceedings of the 17th ACM SIGKDD International Conference on Knowledge Discovery and Data Mining*, KDD '11, pages 316–324, New York, NY, USA, 2011. ACM.

[16] Yu Zheng. T-drive trajectory data sample. https://www.microsoft.com/en-us/research/publication/t-drive-trajectory-data-sample/, August 2011.

# Figures

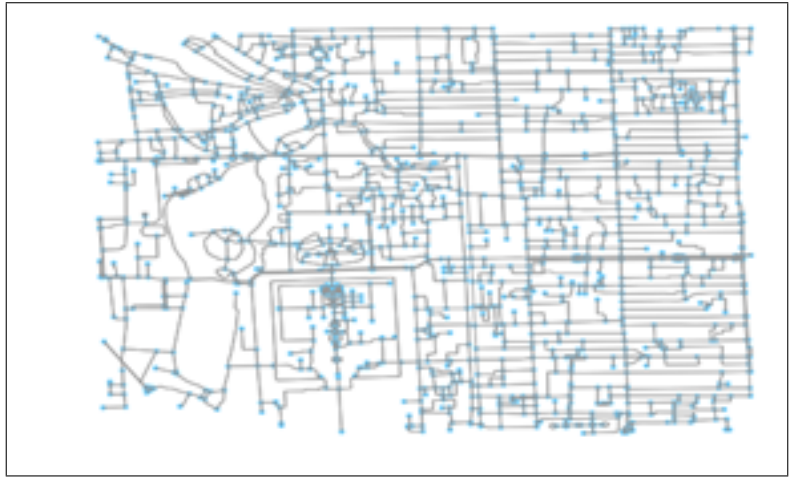

Figure 1: OSMnx graph for the downtown-area of Beijing

```
{10012: [[30888192,146612995,25006941,342687925,190
375162,1903751,175388762,146612956,169587517],
[30888192,169707874,146612995,25006941,169707825,34
2687925,190375162,190375161,175388762,146612956,169
587517],...]...}
```

Figure 2: An example of a single car (with ID 10012) and its assigned routes, split into segments.

```
[[-2918  5864    12 ...,     0     0     0]
 [    0 -2908     8 ...,     0     0     0]
 [    0     0 -2920 ...,     0     0     0]
 ...,
 [    0     0     0 ..., -2925  5854  5854]
 [    0     0     0 ...,     0 -2924  5856]
 [    0     0     0 ...,     0     0 -2924]]
```

Figure 3: QUBO matrix describing the traffic flow problem.

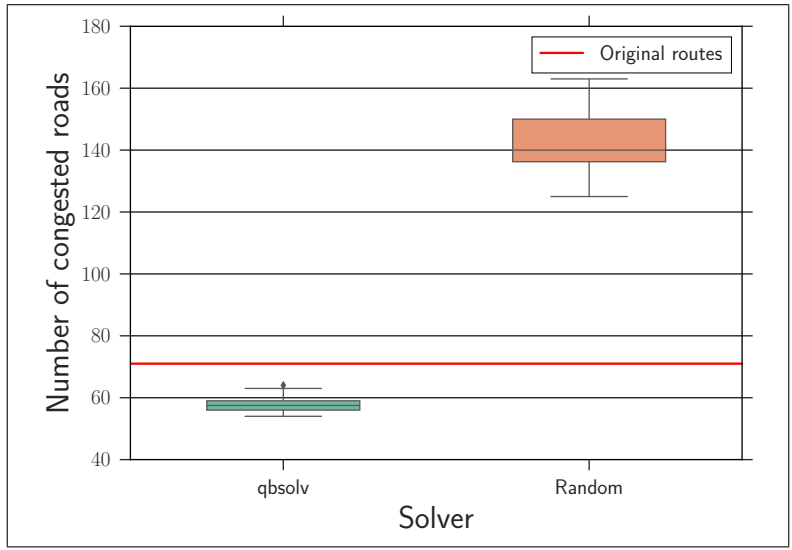

Figure 4: Results comparing random assignment of cars to routes, and `qbsolv` with calls to the D-Wave 2X QPU. The y-axis shows the distribution of number of congested roads. The red line is the number of congested roads given the original assignments of routes.

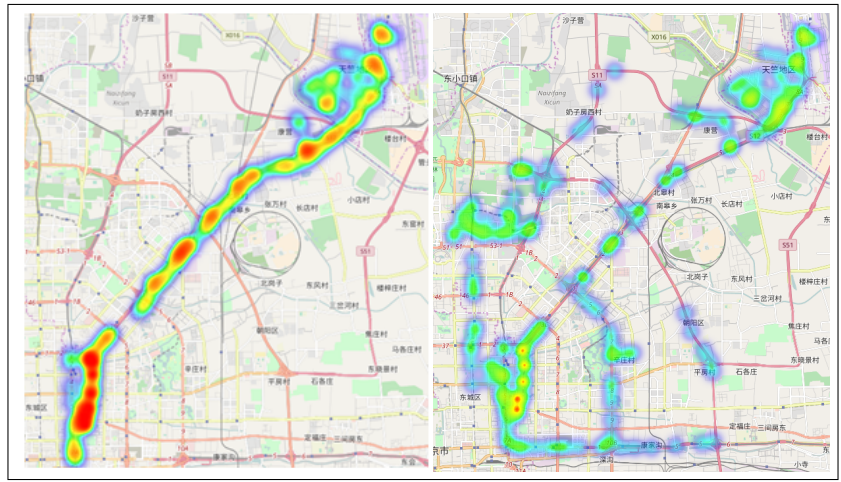

Figure 5: Left: Unoptimized situation under consideration of cars causing traffic jam in the network. Right: Optimized re-distributed cars using `qbsolv`. Note that the areas in red, which indicate high traffic density, are mostly absent from the right picture.